# The arctic curve of the domain-wall six-vertex model in its anti-ferroelectric regime

F. Colomo, A. G. Pronko, and P. Zinn-Justin

ABSTRACT. An explicit expression for the spatial curve separating the region of ferroelectric order ('frozen' zone) from the disordered one ('temperate' zone) in the six-vertex model with domain wall boundary conditions in its anti-ferroelectric regime is obtained.

## 1. Introduction

The six-vertex model with domain wall boundary conditions, introduced in [1] and solved in [2,3], has attracted much attention over the last years, in particular, as an example of a system exhibiting (in an appropriate scaling limit) spatial phase separation phenomena. The model can be regarded as a nontrivial generalization of the famous domino tiling problem of the Aztec diamond, where the celebrated Arctic Circle phenomenon was discovered [4]. Among the many questions concerning this kind of effects, the shape of the curves separating phases is of prime interest [5–10].

Recently, in the series of papers [11–14], a certain progress has been achieved in finding an analytical expression for the curve separating the ferroelectrically ordered (or 'frozen') region from the disordered (or 'temperate') one. By analogy with the domino tilings, this curve is called arctic curve. It is to be mentioned that in the anti-ferroelectric regime one more phase co-exists, and, correspondingly, another separation curve emerges, between the regions of disorder and anti-ferroelectric order [15, 16]. Here we discuss only the arctic curve, for which an analytic expression will be provided.

As shown in [13, 14], the arctic curve can be obtained in the parametric form $x = x(z)$, $y = y(z)$, with $z \in [1, \infty)$, as the solution of the system of equations

$$F(z) = 0, \qquad F'(z) = 0. \tag{1.1}$$

Function $F(z) = F(z; x, y)$ depends on $x$ and $y$ linearly,

$$F(z) = \frac{y}{z-1} - \frac{1-x}{z} - \frac{yt^2}{t^2 z - 2\Delta t + 1} + \lim_{N \to \infty} \frac{\left(\log h_N(z)\right)'}{N} \tag{1.2}$$

where

$$\Delta = \frac{a^2 + b^2 - c^2}{2ab}, \qquad t = \frac{b}{a}. \tag{1.3}$$

and $a$, $b$, and $c$ are the standard weights of the six-vertex model. The function $h_N(z) = h_N(z; \Delta, t)$ appearing in the last term in (1.2) is the generating function for a certain boundary correlation function of the model on an $N$-by-$N$ lattice. By construction, the solution of (1.1) provides only one of the four portions of the arctic curve, limited by $x, y \in [0, \kappa]$, where $\kappa \in (0, 1)$ is the value of the contact point (we assume that the lattice is scaled to the unit square, $[0, 1] \times [0, 1]$); the remaining three portions can be easily obtained by exploiting the symmetries of the model (see [14] for further details).

Derivation of formulae (1.1) and (1.2) exploits a multiple integral representation for certain correlation function (the so-called emptiness formation probability) devised to discriminate ferroelectric order and disorder, and is based on a strongly supported conjecture on the correspondence between condensation of almost all roots of the saddle-point equations and the arctic curve [11–14].

Evidently, to find an explicit formula for the arctic curve one needs to evaluate the last term in (1.2). In [14], this term was worked out for the disordered regime of the model, by extending the technique of paper [17]. However, that approach is unapplicable to the anti-ferroelectric regime. Here, in order to obtain the thermodynamic limit of function $h_N(z)$, we apply the random matrix model technique of paper [18]. This allows us to find an explicit form for the arctic curve of the domain-wall six-vertex model in the anti-ferroelectric regime as well.

## 2. Izergin–Korepin formula and function $h_N(z)$

We recall that the anti-ferroelectric regime corresponds to $\Delta < -1$; the parameter $t$ is arbitrary nonnegative. A convenient parametrization of the weights in terms of the crossing parameter $\eta$ and the rapidity variable $\lambda$ in this regime is

$$a = \sinh(\eta - \lambda), \qquad b = \sinh(\eta + \lambda), \qquad c = \sinh 2\eta. \tag{2.1}$$

For convenience, we mention that this parameterization can be obtained from the one used in [14] (suitable for the disordered regime, see also the appendix) simply by changing $\eta \mapsto \frac{\pi}{2} + i\eta$ and $\lambda \mapsto \frac{\pi}{2} - i\lambda$. The weights in (2.1) are real and positive for $\eta$ and $\lambda$ chosen real, and obeying $\eta > 0$ and $-\eta \leqslant \lambda \leqslant \eta$.

The Izergin–Korepin formula describes the partition function of the inhomogeneous six-vertex model with domain wall boundary conditions. Recall that the model is considered on the square lattice formed by intersection of $N$ vertical and $N$ horizontal lines. Introduce two sets of rapidity variables $\{\lambda_j\}_{j=1}^N$ and $\{\nu_k\}_{k=1}^N$, and let $a_{jk}$, $b_{jk}$, and $c_{jk}$ denote weights of the vertex being at intersection of $k$-th horizontal and $j$-th vertical lines, obtained by replacing $\lambda \mapsto \lambda_j - \nu_k$ in (2.1). The partition function of the domain-wall six-vertex model with these weights can be represented in terms of an $N$-by-$N$ determinant:

$$Z_N(\lambda_1,\ldots,\lambda_N;\nu_1,\ldots,\nu_N) = \frac{\prod_{j,k=1}^{N}\sinh(\eta-\lambda_j+\nu_k)\sinh(\eta+\lambda_j-\nu_k)}{\prod_{1\leq j<k\leq N}\sinh(\lambda_k-\lambda_j)\sinh(\nu_j-\nu_k)} \\ \times \det_{1\leq j,k\leq N}\left[\varphi(\lambda_j-\nu_k)\right], \qquad (2.2)$$

where

$$\varphi(\lambda) := \frac{\sinh 2\eta}{\sinh(\eta-\lambda)\sinh(\eta+\lambda)}. \qquad (2.3)$$

For a proof see [2, 3]; another proof can be found in [11, 19].

A formula for the partition function of the homogeneous model can be found by evaluating the 'homogenous limit' in the Izergin–Korepin formula, namely, $\lambda_1,\ldots,\lambda_N \to \lambda$ and $\nu_1,\ldots,\nu_N \to 0$. The partition function reads

$$Z_N(\lambda,\ldots,\lambda;0,\ldots,0) = \frac{[\sinh(\eta-\lambda)\sinh(\eta+\lambda)]^{N^2}}{\prod_{j=1}^{N-1}(j!)^2} \det_{1\leq j,k\leq N}\left[\partial_\lambda^{j+k-2}\varphi(\lambda)\right]. \qquad (2.4)$$

The derivation simply uses Taylor expansion of entries of the rows and columns of the determinant [2, 3].

Now we are ready to turn to the function $h_N(z)$ entering the equations for the arctic curve. This function is defined as the generating function

$$h_N(z) = \sum_{r=1}^{N} H_N^{(r)} z^{r-1}. \qquad (2.5)$$

Here $H_N^{(r)} = H_N^{(r)}(\lambda,\eta)$ is the one-point 'boundary' correlation function which gives the probability of having the sole $c$-weight vertex of the boundary row at the $r$th position. It describes a peculiarity of the configurations of the domain-wall six-vertex model, namely, that a boundary row always contains a string of $a$-weight vertices, next a single $c$-weight vertex, and finally a string of $b$-weight vertices.

The correlation function $H_N^{(r)}$ admits two similar but actually different representations. In [19] it was computed as a determinant analogous to that in (2.4) but with the last column modified; this representation played an important role in the derivation of the arctic curve in [11–14].

Another representation, which is crucial for us below, is based on the aforementioned peculiarity of the configurations, and relates this correlation function with the Izergin–Korepin formula with one inhomogeneity [20]. Consider an incomplete homogeneous limit in which all rapidity variables but one, say $\lambda_1$, tend to their homogeneous limit values, $\lambda_2,\ldots,\lambda_N \to \lambda$ and $\nu_1,\ldots,\nu_N \to 0$. Setting $\lambda_1 = \lambda + \xi$, where $\xi$ is a new variable, the connection with the Izergin–Korepin formula, when rephrased in terms of generating function (2.5), reads (see, e.g., appendix A in [14]):

$$h_N(\gamma(\xi)) = \left[\frac{\sinh(\eta - \lambda)}{\sinh(\eta - \lambda - \xi)}\right]^{N-1} \frac{Z_N(\lambda + \xi, \lambda, \ldots, \lambda; 0, \ldots, 0)}{Z_N(\lambda, \ldots, \lambda; 0, \ldots, 0)}. \quad (2.6)$$

Here the function $\gamma(\xi)$ is given by

$$\gamma(\xi) := \frac{\sinh(\eta - \lambda)}{\sinh(\eta + \lambda)} \frac{\sinh(\eta + \lambda + \xi)}{\sinh(\eta - \lambda - \xi)}. \quad (2.7)$$

Hence, the last term in (1.2) can be found by evaluating the thermodynamic limit of Izergin–Korepin formula with one inhomogeneity.

## 3. Random matrix model formulation

In [18] the methods of random matrix models were applied to evaluate the thermodynamic limit of the partition function (2.4). Here we consider an adaptation of the same technique to evaluate in this limit the function (2.6).

The starting point of the method is to consider the Laplace transform for the function $\varphi(\lambda)$,

$$\varphi(\lambda) = \int e^{\lambda z} \, dm(z), \quad (3.1)$$

where $dm(z) = dm(z; \lambda, \eta)$ is some measure, which is in the case of the anti-ferroelectric regime is a discrete one, and whose explicit form was provided in [18]. Let us define the quantity

$$I_N(\lambda_1, \ldots, \lambda_N; \nu_1, \ldots, \nu_N) = \frac{\det_{1 \leq j,k \leq N} [\varphi(\lambda_j - \nu_k)]}{\prod_{1 \leq j < k \leq N} \sinh(\lambda_k - \lambda_j) \sinh(\nu_j - \nu_k)}. \quad (3.2)$$

Using the properties of the determinant and the symmetry of the $N$-fold integration measure, we can rewrite this expression in the form

$$I_N(\lambda_1, \ldots, \lambda_N; \nu_1, \ldots, \nu_N) = \frac{1}{\prod_{1 \leq j < k \leq N} \sinh(\lambda_k - \lambda_j) \sinh(\nu_j - \nu_k)}$$
$$\times \frac{1}{N!} \int \det_{1 \leq j,k \leq N} \left[e^{\lambda_j z_k}\right] \det_{1 \leq j,k \leq N} \left[e^{-\nu_j z_k}\right] dm(z_1) \cdots dm(z_N), \quad (3.3)$$

which can be viewed as the partition function of a matrix model in a 'double external field' (see section 2.5.4 of [21]).

Let us consider the homogeneous limit, namely, $\lambda_1, \ldots, \lambda_N \to \lambda$ and $\nu_1, \ldots, \nu_N \to 0$ of the expression above. For example, evaluating the limit as $\lambda_j$'s we get

$$\lim_{\lambda_1, \ldots, \lambda_N \to \lambda} \frac{\det_{1 \leq j,k \leq N} \left[e^{\lambda_j z_k}\right]}{\prod_{1 \leq j < k \leq N} \sinh(\lambda_k - \lambda_j)} = \frac{e^{\lambda(z_1 + \cdots + z_N)}}{\prod_{j=1}^{N-1} j!} \prod_{1 \leq j < k \leq N} (z_k - z_j), \quad (3.4)$$

and an essentially similar expression in evaluating the limit in $\nu_k$'s. Denoting $I_N := I_N(\lambda, \ldots, \lambda; 0, \ldots, 0)$, we readily obtain

$$I_N = \frac{1}{N! \prod_{j=1}^{N-1}(j!)^2} \int dm_N(\{z_j\}), \qquad (3.5)$$

with

$$dm_N(\{z_j\}) := e^{\lambda(z_1 + \cdots + z_N)} \prod_{1 \leqslant j < k \leqslant N} (z_k - z_j)^2 \, dm(z_1) \ldots dm(z_N), \quad (3.6)$$

which is in fact the one-matrix model expression of paper [18] for the partition function (2.4).

Let us now consider the case when $\lambda_2, \ldots, \lambda_N \to \lambda$, but $\lambda_1 = \lambda + \xi$, where $\xi$ is arbitrary. In this case one can first expand the determinant appearing in the numerator along its first column, and next evaluate the limit

$$\lim_{\lambda_2, \ldots, \lambda_N \to \lambda} \frac{\det_{1 \leqslant j,k \leqslant N}\left[e^{\lambda_j z_k}\right]}{\prod_{1 \leqslant j < k \leqslant N} \sinh(\lambda_k - \lambda_j)}$$
$$= \frac{(-1)^{N-1} e^{\lambda(z_1 + \cdots + z_N)}}{(\sinh \xi)^{N-1} \prod_{j=1}^{N-2} j!} \sum_{l=1}^{N} (-1)^{l-1} e^{\xi z_l} \prod_{\substack{1 \leqslant j < k \leqslant N \\ j,k \neq l}} (z_k - z_j), \quad (3.7)$$

replacing further the sum with a contour integral (see, e.g., [22]),

$$\sum_{l=1}^{N} (-1)^{N-l} e^{\xi z_l} \prod_{\substack{1 \leqslant j < k \leqslant N \\ j,k \neq l}} (z_k - z_j) = \prod_{1 \leqslant j < k \leqslant N} (z_k - z_j) \oint_C \frac{e^{\xi z}}{\prod_{j=1}^{N}(z - z_j)} \frac{dz}{2\pi i}, \qquad (3.8)$$

where $C$ is a simple closed counterclockwise-oriented contour enclosing all $z_j$'s. As a result, for the quantity $I_N(\xi) := I_N(\lambda + \xi, \lambda, \ldots, \lambda; 0, \ldots, 0)$ we obtain

$$I_N(\xi) \propto \frac{1}{(\sinh \xi)^{N-1}} \int dm_N(\{z_j\}) \oint_C \frac{e^{\xi z}}{\prod_{j=1}^{N}(z - z_j)} \, dz. \qquad (3.9)$$

Here and below we neglect, for simplicity, all factors not contributing to the logarithmic derivative of $I_N(\xi)$, which is the actual quantity we are interested in (see equations (1.2) and (2.6)).

When considering the large $N$ limit of expression of the type appearing in (3.5) and (3.9) in the framework of the saddle-point approximation, experience from random matrix models suggests to rescale the 'eigenvalues' $z_j$, by a factor $N$, namely, $z_j \mapsto N z_j$. Indeed, to have a non-trivial saddle-point all terms in the logarithm of the integrand have to be of the same order. The solution of the saddle-point equation is encoded in the resolvent $W(z)$, defined as

$$W(z) := \lim_{N \to \infty} \frac{1}{N} \sum_{j=1}^{N} \frac{1}{z - z_j} \qquad (3.10)$$

where now the $z_j$'s are the solution of the set of saddle-point equations associated to multiple integral (3.5).

In the case of expression (3.9), rescaling the 'eigenvalues' $z_j \mapsto N z_j$, and simultaneously replacing $z \mapsto Nz$, we get

$$I_N(\xi) \propto \frac{1}{(\sinh \xi)^{N-1}} \int d\tilde{m}_N(\{z_j\}) \oint_C \frac{e^{N\xi z}}{\prod_{j=1}^N (z - z_j)} \, dz, \qquad (3.11)$$

where $\tilde{m}_N(\{z_j\}) := m_N(\{Nz_j\})$. The crucial point is that the set of saddle-point equations relative to $z_j$'s remains the same as for the case of $I_N$, and the corresponding solution is still encoded in $W(z)$. We have however one additional saddle-point equation, relative to variable $z$:

$$\xi = W(z_{\text{s.p.}}). \qquad (3.12)$$

In other words, the saddle-point for the extra integration variable is simply given by the functional inverse of the resolvent, evaluated at $\xi$:

$$z_{\text{s.p.}} = W^{-1}(\xi). \qquad (3.13)$$

Inversion relation (3.13) appears in multiple settings; see related work on the HCIZ integral [23] and on the connection with the R-transform in free probability theory [24].

Differentiating expression (3.11) with respect to $\xi$ we find

$$\frac{1}{N}(\log I_N(\xi))' = -\frac{N-1}{N} \coth \xi + \frac{\int d\tilde{m}_N(\{z_j\}) \oint \frac{\exp(N\xi z)}{\prod_{j=1}^N (z-z_j)} z \, dz}{\int d\tilde{m}_N(\{z_j\}) \oint \frac{\exp(N\xi z)}{\prod_{j=1}^N (z-z_j)} \, dz} \qquad (3.14)$$

and hence in the large $N$ limit we obtain

$$\lim_{N \to \infty} \frac{1}{N}(\log I_N(\xi))' = -\coth \xi + z_{\text{s.p.}}. \qquad (3.15)$$

Recalling equation (2.6) and inversion relation (3.13) we have

$$\lim_{N \to \infty} \frac{1}{N} \frac{\partial}{\partial \xi} \log h_N(\gamma(\xi)) = \coth(\xi + \lambda + \eta) - \coth \xi + W^{-1}(\xi). \qquad (3.16)$$

Thus the only input needed for the explicit calculation of the last term in (1.2) is the resolvent $W(z)$. This has been computed explicitly both for disordered and anti-ferroelectric regimes of the domain-wall six-vertex model in [18] (see also [25, 26]). In appendix we sketch the inversion of the resolvent for the disordered regime, showing that the result of paper [14] is indeed reproduced by the present method.

## 4. The arctic curve in the anti-ferroelectric regime

In the anti-ferroelectric regime the large $N$ solution of the random matrix formulation of $I_N$ gives rise to a resolvent with two cuts, separated by a Douglas–Kazakov type saturated region. Recasting the result for the resolvent of papers [18, 26] into our notations, we have

$$W(z) = -\frac{\eta}{K}[u(\eta z) - u_\infty], \tag{4.1}$$

where the function $u(z)$ is defined in terms of Jacobian elliptic functions

$$\operatorname{sn}^2 u(z) := \frac{\beta - z}{\beta' - z} \operatorname{sn}^2 u_\infty, \tag{4.2}$$

with elliptic nome $q = \exp(-\pi^2/2\eta)$, and $K$ is the corresponding complete elliptic integral of the first kind. Quantities $\beta$, $\beta'$, and $u_\infty := u(z)\big|_{z=\infty}$ are given by

$$\beta = \pi\frac{\vartheta_1'(\varkappa)}{\vartheta_1(\varkappa)}, \qquad \beta' = \pi\frac{\vartheta_4'(\varkappa)}{\vartheta_4(\varkappa)}, \qquad u_\infty = \frac{2K}{\pi}\varkappa, \tag{4.3}$$

where

$$\varkappa := \pi\frac{\eta - \lambda}{4\eta}, \tag{4.4}$$

and we also recall that in the anti-ferroelectric regime $\eta > 0$ and $-\eta < \lambda < \eta$, so that $\varkappa \in [0, \pi/2]$.

Functional relation (4.1) can be readily inverted with the result

$$z = W^{-1}(\xi) = \frac{\beta \operatorname{sn}^2 u_\infty - \beta' \operatorname{sn}^2(u_\infty - K\xi/\eta)}{\eta \operatorname{sn}^2 u_\infty - \eta \operatorname{sn}^2(u_\infty - K\xi/\eta)}. \tag{4.5}$$

Denoting

$$\alpha := \frac{\pi}{2\eta} \tag{4.6}$$

and substituting the values of the constants $\beta$, $\beta'$, and $u_\infty$ given above, and switching to theta functions, we find

$$\begin{aligned}
W^{-1}(\xi) &= 2\alpha\frac{\vartheta_1(\varkappa)\vartheta_1'(\varkappa)\vartheta_4^2(\varkappa - \alpha\xi) - \vartheta_4(\varkappa)\vartheta_4'(\varkappa)\vartheta_1^2(\varkappa - \alpha\xi)}{\vartheta_1^2(\varkappa)\vartheta_4^2(\varkappa - \alpha\xi) - \vartheta_4^2(\varkappa)\vartheta_1^2(\varkappa - \alpha\xi)} \\
&= \alpha\frac{\vartheta_1'(2\varkappa - \alpha\xi)\vartheta_1(\alpha\xi) - \vartheta_1(2\varkappa - \alpha\xi)\vartheta_1'(\alpha\xi)}{\vartheta_1(2\varkappa - \alpha\xi)\vartheta_1(\alpha\xi)} \\
&= \alpha\frac{\vartheta_1'(\alpha\xi)}{\vartheta_1(\alpha\xi)} - \alpha\frac{\vartheta_1'(\alpha(\xi + \lambda + \eta))}{\vartheta_1(\alpha(\xi + \lambda + \eta))}.
\end{aligned} \tag{4.7}$$

Plugging this expression into (3.16) one obtains an explicit form of the logarithmic derivative of the thermodynamic limit of function $h_N(\gamma(\xi))$. Note moreover its striking similarity with the corresponding expression for the disordered regime, see (A.7).

Now we are ready to write an explicit formula for the arctic curve in the anti-ferroelectric regime. As in [14], we switch from function $F(z)$, given by (1.2), to function $f(\xi)$, defining it as $f(\xi) := \gamma'(\xi)F(\gamma(\xi))$. Direct calculation gives

$$f(\xi) = x\varphi(\xi + \lambda) + y\varphi(\xi - \eta) - \Psi(\xi) \tag{4.8}$$

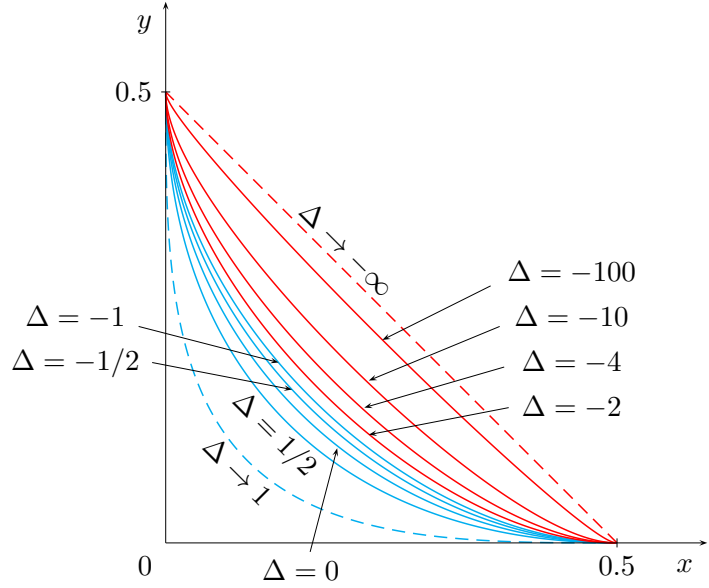

FIGURE 1. The arctic curve of the domain-wall six-vertex model for various values of $\Delta$, at $t=1$.

where function $\varphi(\lambda)$ is given by (2.3) and function $\Psi(\xi)$ reads

$$\Psi(\xi) := \varphi(\xi + \lambda) - \lim_{N\to\infty} \frac{1}{N} \frac{\partial}{\partial \xi} \log h_N(\gamma(\xi))$$
$$= \coth \xi - \coth(\xi + \lambda - \eta) - \alpha \frac{\vartheta_1'(\alpha\xi)}{\vartheta_1(\alpha\xi)} + \alpha \frac{\vartheta_1'(\alpha(\xi + \lambda + \eta))}{\vartheta_1(\alpha(\xi + \lambda + \eta))}. \quad (4.9)$$

Note that varying $z$ over the interval $[1, \infty)$ corresponds to $\xi \in [0, \eta - \lambda]$ and since function $\gamma(\xi)$ is monotonously increasing over this interval we have $\gamma'(\xi) \neq 0$ that allows us to describe the arctic curve in terms of function $f(\xi)$ instead of $F(z)$.

Equations (1.1) are replaced by the equations

$$f(\xi) = 0, \qquad f'(\xi) = 0, \quad (4.10)$$

and solving this linear system for $x$ and $y$ we obtain:

$$x = \frac{\varphi'(\xi - \eta)\Psi(\xi) - \varphi(\xi - \eta)\Psi'(\xi)}{\varphi(\xi + \lambda)\varphi'(\xi - \eta) - \varphi(\xi - \eta)\varphi'(\xi + \lambda)},$$
$$y = \frac{\varphi(\xi + \lambda)\Psi'(\xi) - \varphi'(\xi + \lambda)\Psi(\xi)}{\varphi(\xi + \lambda)\varphi'(\xi - \eta) - \varphi(\xi - \eta)\varphi'(\xi + \lambda)}. \quad (4.11)$$

We recall that here functions $\varphi(\lambda)$ and $\Psi(\xi)$ are given by (2.3) and (4.9), respectively.

Formulae (4.11) give one portion (out of four) of the arctic curve in a parametric form, with $\xi$ being the parameter of the curve, $\xi \in [0, \eta - \lambda]$. The value $\xi = 0$ corresponds to the contact point of the arctic curve with the $x$-axis, and as $\xi$ increases, the whole curve is monotonously constructed, up to the contact point with the $y$-axis, at $\xi = \eta - \lambda$.

The remaining three portions of the arctic curve can be easily constructed using symmetry considerations, along the lines given in [14]. The only modification, specific of the anti-ferroelectric regime, concerns the crossing symmetry transformation (which involve exchange of the weights $a \leftrightarrow b$), which for the parameterization of weights (2.1) requires changing the sign of $\lambda$, $\lambda \mapsto -\lambda$.

In figure 1 we plot (one of the four portions of) the arctic curve of the model for several values of $\Delta$, restricting for simplicity to the case of $t = 1$ (i.e., when weights obey $a = b$). For $-1 \leqslant \Delta < 1$ (disordered regime) we plot the expression for the arctic curve given in [14], while for $\Delta < -1$ (anti-ferroelectric regime) we are plotting expression (4.11) above. In the limiting case of $\Delta \to -\infty$ the arctic curve can be easily computed by other means (see, for instance, [18]). It is apparent from the figure, and can be easily checked analytically, that the general expression (4.11) for the arctic curve indeed reproduces in this case the expected result, i.e., a straight line segment. This is an additional indication for the validity of the 'condensation hypothesis' of papers [13,14], which is at the basis of formulae (1.1) and (1.2).

## Acknowledgments

We thank the organizers of the trimester "Statistical physics, combinatorics and probability: from discrete to continuous models", Autumn 2009, on the occasion of which part of the present work has been done. FC thanks the Centre Émile Borel, Institut Henri Poincaré for its kind hospitality. FC acknowledges partial support from MIUR, PRIN grant 2007JHLPEZ, and from the European Science Foundation program INSTANS. AGP is supported by the Alexander von Humboldt Foundation. AGP also acknowledges partial support from INFN, Sezione di Firenze, and from the Russian Academy of Sciences program "Mathematical Methods in Nonlinear Dynamics". PZJ acknowledges partial support from ESF program "MISGAM" and ANR program "GRANMA" BLAN08-1-13695.

## Appendix A. Inverse resolvent for the disordered regime

Here we sketch the derivation for the case of the disordered regime, showing that indeed the result of [14] for the last term in (1.2) is reproduced through the inversion of the resolvent.

The disordered regime corresponds to $-1 \leqslant \Delta < 1$ and the weights can be parameterized as follows

$$a = \sin(\lambda + \eta), \qquad b = \sin(\lambda - \eta), \qquad c = \sin 2\eta, \qquad (A.1)$$

where $\lambda$ and $\eta$ are real, and obey the conditions $0 < \eta \leqslant \pi/2$ and $\eta \leqslant \lambda \leqslant \pi - \eta$. Function $\varphi(\lambda)$ is defined as

$$\varphi(\lambda) = \frac{\sin 2\eta}{\sin(\lambda+\eta)\sin(\lambda-\eta)}. \qquad (A.2)$$

Its Laplace transform provides an integral representation which leads to the random matrix model formulation, now with a continuous measure with support on the real axis [18].

Repeating the procedure of Section 3 leads to the following formula for the last term in (1.1) in terms of the resolvent

$$\lim_{N\to\infty} \frac{1}{N}\frac{\partial}{\partial \xi}\log h_N(\gamma(\xi)) = \cot(\xi+\lambda-\eta) - \cot\xi + W^{-1}(\xi). \qquad (A.3)$$

The large $N$ solution of the random matrix formulation of $I_N$ in the disordered regime gives rise to a single cut resolvent (see [18, 25]), given explicitly by

$$W(z) = \frac{\varkappa}{\alpha} + \frac{1}{i\alpha}\log\left[\frac{\sqrt{z+2\alpha\tan\varkappa} - i\sqrt{z\tan^2\varkappa - 2\alpha\tan\varkappa}}{\sqrt{z(\tan^2\varkappa+1)}}\right], \qquad (A.4)$$

where now

$$\varkappa := \frac{\pi(\pi-\eta-\lambda)}{2(\pi-2\eta)}, \qquad \alpha := \frac{\pi}{\pi-2\eta}. \qquad (A.5)$$

The functional relation $W(z) = \xi$ can be easily inverted,

$$z = W^{-1}(\xi) = -2\alpha\tan\varkappa \frac{\tan^2(\varkappa-\alpha\xi)+1}{\tan^2(\varkappa-\alpha\xi)-\tan^2\varkappa}. \qquad (A.6)$$

An elementary calculation leads to the formula

$$W^{-1}(\xi) = \alpha\cot\alpha\xi - \alpha\cot\alpha(\xi+\lambda-\eta) \qquad (A.7)$$

which indeed reproduces the result obtained by other methods in [14].

INFN, Sezione di Firenze, Via G. Sansone 1, 50019 Sesto Fiorentino (FI), Italy
  *E-mail address*: `colomo@fi.infn.it`

Fachbereich C – Physik, Bergische Universität Wuppertal, 42097 Wuppertal, Germany (*On leave of absence from*: Saint Petersburg Department of V. A. Steklov Mathematical Institute, Russian Academy of Sciences, Fontanka 27, 191023 Saint Petersburg, Russia)
  *E-mail address*: `agp@pdmi.ras.ru`

UPMC Univ Paris 6, CNRS UMR 7589, LPTHE, 75252 Paris Cedex, France
  *E-mail address*: `pzinn@lpthe.jussieu.fr`